# Possibilities of MgB$_2$/Cu Wires Fabricated by the *in-situ* Reaction Technique

E. Martínez* and R. Navarro

*Instituto de Ciencia de Materiales de Aragón (CSIC - Universidad de Zaragoza)*
*Departamento de Ciencia y Tecnología de Materiales y Fluidos*
CPS*, C/ María de Luna 3, 50018 Zaragoza, Spain*

**Abstract**— The superconducting properties of copper-sheathed MgB$_2$ wires fabricated by conventional powder-in-tube techniques and the *in-situ* reaction procedure are analysed. The influence of the processing conditions and initial $(1+x)$Mg + 2B ($x = 0$, 0.1, 0.2) proportions of the precursors on the critical current values of the wires have been studied. In particular, the limits of the available temperatures and times for heat treatments imposed by the chemical reaction between Mg and Cu, and their effect on the superconducting properties of the wires, are discussed. The analysis includes the study of the sample microstructure and phase composition as well as of the critical current temperature and field dependences. The wires show high thermal stability during direct transport measurements and carry a critical current density of $1.3 \times 10^9$ A/m$^2$ at 15 K in the self-field for optimised processing conditions.

* E-mail: elenamar@unizar.es

---

## 1. Introduction

The discovery of MgB$_2$ superconductor [1] has opened great expectations as it has been revelled as a suitable candidate for practical large-scale applications. Key properties of such interest are: *a)* the relatively high critical temperature, $T_c \sim 39$ K, allowing working temperatures on the range of 20-30 K, nowadays easily reachable with cryocoolers; *b)* the low cost and rather easy preparation of the rough materials; and *c)* the good grain connectivity and the absence of week links [2].

Fortunately, powder-in-tube (PIT) technologies, widely searched at laboratory and fully developed at industrial scale in recent years for the fabrication of PbBi-2223 tapes, have been proved to be useful in the conformation of MgB$_2$/metal composite wires. This is, indeed, one of the most attractive technologies for long-wire fabrication of hard and brittle materials because of its potential scalability and production flexibility. In the PIT methods, metallic tubes, with the precursor powders packed inside, are drawn to wires or rolled to tapes. The precursors may be mixtures of unreacted Mg and B powders, the so called *in-situ* reaction approach [3-6]; pre-reacted MgB$_2$ powders [4,5,7-10], called *ex-situ* reaction technique; or partially reacted methods containing mixtures of MgB$_2$, Mg and B powders [5,11]. In most cases, heat treatments in absence of oxygen are used either to sinter or to react the precursor. Although unsintered conductors made from reacted MgB$_2$ powders already have appreciable $J_c$ values up to $10^9$ Am$^{-2}$ at 4.2 K in the self-field, the magnetic field dependence is improved notably by sintering [4,8-10].

All these methods have their own advantages and drawbacks. Thus, *ex-situ* procedures usually yield wires and tapes with smaller grains and more homogeneous microstructures, resulting in improved $J_c(B)$ dependences, probably because of grain boundary pinning. On the contrary, it has the drawback of being very sensitive to the $MgB_2$ precursor impurities [12]. On the other hand, the *in-situ* procedure uses lower heat treatment temperatures, allowing a larger range of metallic sheaths, being then more flexible in the material selection. Nevertheless, since the density of the initial Mg+2B mixtures are significantly lower than that of the $MgB_2$ phase, this method has the intrinsic disadvantage of a low final density.

The selection of adequate sheaths, giving thermal, electrical and mechanical stability without deterioration of the superconductor, is crucial and constitutes nowadays an open issue to be addressed prior to reach technologically useful composite $MgB_2$/metal conductors. Different metals sheaths have already been used: iron [5, 10], copper [3,6,7], nickel [7,8,10], silver [3,6,7], cupro nickel alloys [9] and stainless steel [9,13], as well as different metal combinations such as: (from inside to outside) Ta/Cu [14] and mechanically enforced Fe/Cu/SS [4]. Up to now the highest $J_c(B,T)$ values are obtained with hard metal sheaths that do not react with Mg or $MgB_2$, such as steel and iron, but these wires, as consequence of a poor thermal stability, easily quench at low fields. On the other hand, silver sheathed wires [3,6,7] present very poor results compared with copper or nickel sheathed wires and therefore have been disregarded.

The use of copper as sheath would be an excellent choice for this purpose due to its high thermal conductivity. Copper is a low cost and ductile metal very suitable for the PIT fabrication of $MgB_2$ composite wires and tapes. Moreover, as in low temperature superconducting wires, it might give good mechanical support and excellent thermal stability. In addition, the good soldability of Cu would facilitate the achievement of low resistance electric contacts to external current feeding sources. Nevertheless, the chemical reaction of Mg and $MgB_2$ with Cu, during the compulsory heat treatments, causes a reduction of the wire $J_c$ values. This reactivity, which is produced by the diffusion of Mg into the Cu, together with a strong reduction of the melting temperature of the produced Cu-Mg alloys, may be avoided using costly and more complex to process diffusions barriers.

Trying to improve the thermal stability of $MgB_2$/metal composite wires, which is a major technological challenge, it is worthwhile a deeper study to determine the limits of the superconducting properties of $MgB_2$/ Cu wires associated to the reactivity of copper and Mg. Cu may be also an interesting material since it alloys well with Ni, which has also been used with quite good results [8], and thus the use of different Cu-Ni proportions would allow the control of thermal, mechanical and electrical properties of the sheath.

With this aim here we present results on PIT monofilamentary wires fabricated from Cu tubes with unreacted mixtures of Mg and B powders using stoichiometric as well as Mg-rich compositions. The influence of the processing conditions and the initial Mg: B proportions on the final wire $J_c$ values have been studied. Special attention is paid to the limits of the thermal treatments temperatures and times imposed by the reactivity of Mg in Cu and their effect on the superconducting properties of the wires. With this aim, a complete analysis of the microstructure and phase compositions of the wires by SEM and EDX is presented, together with measurements of the temperature and field dependences of the critical currents, obtained directly by transport and from the *M-B* hysteresis loops.

## 2. Experimental details

Single filament Cu-sheathed $MgB_2$ wires were prepared using the conventional PIT method. The initial powder was a mixture of Mg (Goodfellow 99.8% purity with grain size lower than 250 μm) and amorphous boron (Alfa - Aesar, 99.99% purity and 355 mesh). Powders of atomic stoichiometry and with excess of Mg were prepared with three different proportions $(1+x)Mg + 2B$, with $x = 0$, 0.1 and 0.2. These precursor powders were well mixed in a vibratory mill for 10 minutes and packed inside copper tubes of 4.0 mm outer diameter and 2.5 or 3.0 mm inner ones. Subsequently, the tubes were cold drawn in round dyes down to 1.1 or 1.2 mm final diameters in 0.1-mm reduction steps. The final wires have core diameters of 0.6 or 0.8 mm, typically, depending on the initial inner diameter of the copper tubes.

The wire was then cut in 6 to 7 cm long pieces and their ends mechanically sealed with soft copper to avoid leaking of Mg during the annealing. To prevent oxidation, the in-wire reaction was done in sealed quartz tubes with argon at temperatures ranging from 620 to 700 °C and times from 15 minutes to 48 h. Fig. 1 shows a typical temperature ramp of the sample heat-treatment followed by slow cooling inside the furnace or room temperature quenching.

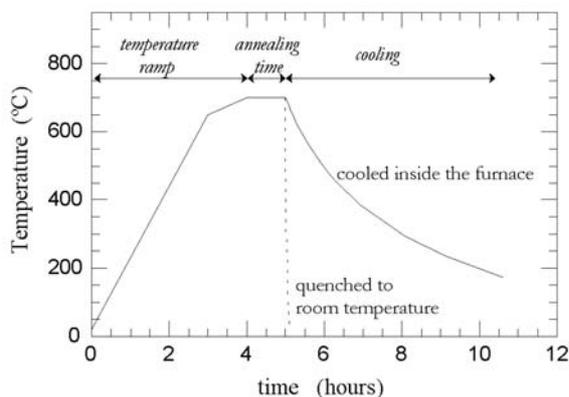

*Figure 1* Typical annealing conditions and cooling rates used in the in-situ fabrication of $MgB_2$ / Cu composite wires.

The microstructure and phase composition of the wires were analysed by SEM and X-ray energy dispersive spectroscopy (EDX) techniques, respectively. A commercial Quantum Design SQUID magnetometer was used to perform AC and DC magnetic measurements over 4 mm long wire samples, keeping their axis perpendicular to the applied field. The critical temperature, $T_c$, and the superconductor-normal transition were analysed by AC susceptibility measurements, $\chi_{ac}(T)$, using AC field with amplitudes of 0.1 mT and low frequency of 1 Hz, to minimise the contribution of the eddy currents induced in the metallic sheaths. Isothermal magnetisation DC loops, $M(B)$, up to fields of 5 T, were also measured.

Electric DC transport measurements were performed on 6 cm long wires at temperatures ranging from 15 to 40 K using a cryocooler system and sinusoidal current ramps of typically 1 to 3 seconds scan time. The critical currents, $I_c(T)$, were determined by the standard four-probe using the 1 μV/cm criterion.

## 3. Results and discussion

### 3.1 Microstructure and reactivity

A typical longitudinal cross-section of $MgB_2$/Cu composite wires after annealing is shown in Figs. 2(a) and 2(b), which correspond to a wire with $x = 0$, annealed at 700 °C during 30 minutes and afterwards cooled down inside the furnace. The superconducting core has a typical irregular shape, partially produced during the mechanical conformation because of the lower hardness of copper and also during the heat treatment by the reactivity of Mg from the core with the inner sheath wall. A reaction layer adjacent to the superconducting core with darker contrast and 20 to 30 μm thickness is easily observed in Fig. 2(b). EDX and X-Ray analyses [6], have indicated that this layer corresponds to $MgCu_2$ while the rest of the sheath remains pure copper. Inside the core there are also some homogeneously dispersed $MgCu_2$ grains, with sizes typically ranging from 20 to 80 μm. Moreover, as point out previously, it should be remarked that all samples show larger porosity than those made from reacted precursor.

The predominant superconducting $MgB_2$ phase, observed by the darker grey contrast in the SEM pictures of Figs. 2(a) and 2(b) have eutectic-like structure [6]. An enlarged view of this structure, which contains 8±2 atm.% of Cu, is show in Fig. 2(c), where the darker grey contrast corresponds to the superconducting $MgB_2$ phase and the lighter phase to a Mg-Cu intermetallic compound, probably $Mg_2Cu$. The formation of these intermetallic compounds causes an overall Mg deficit, which may induce the formation of boron-rich phases at given points (darker areas in Fig. 2(d)). This is more likely on samples annealed at lower temperatures and longer times, while is not observed in samples with Mg excess ($x = 0.2$).

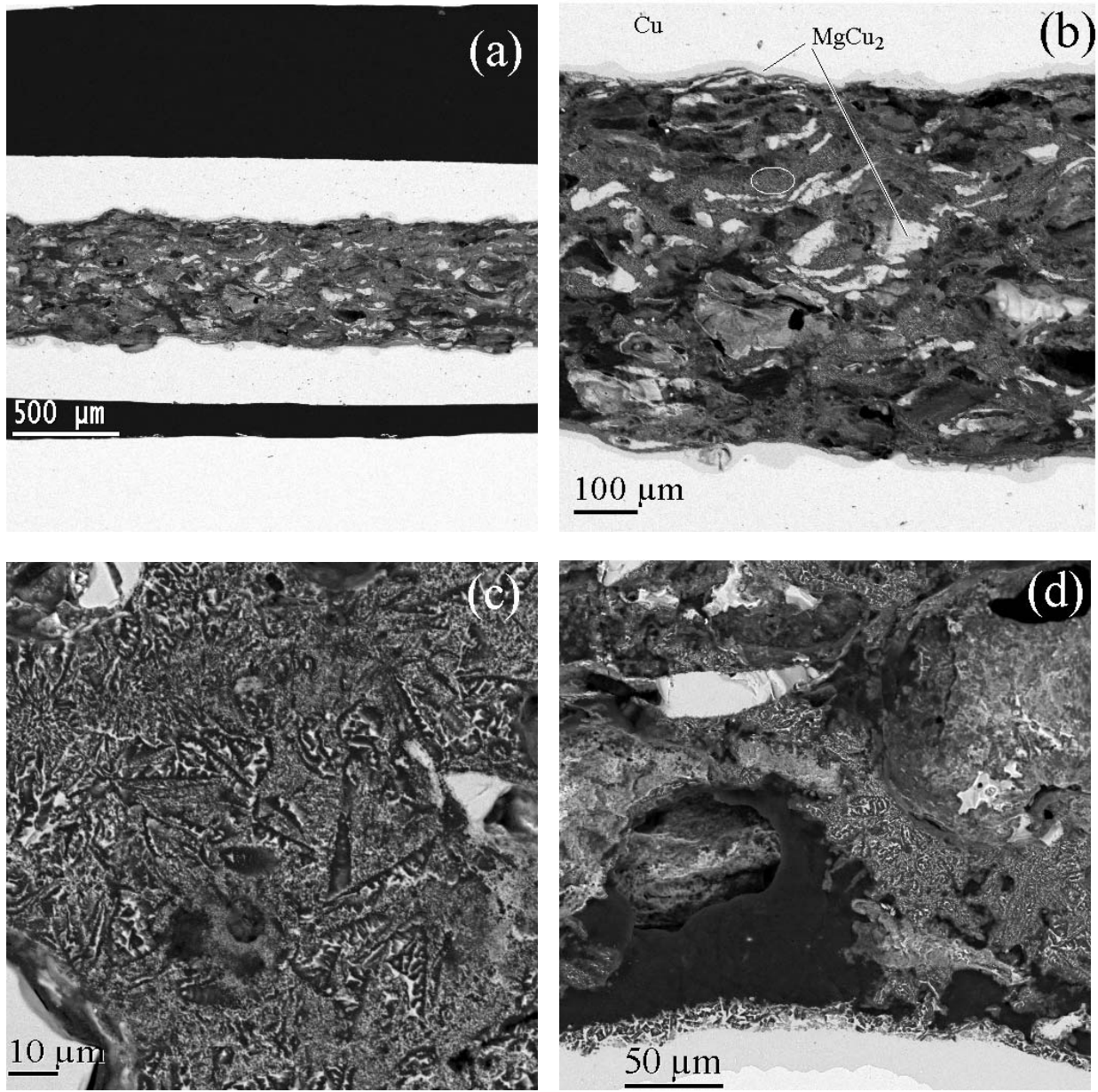

*Figure 2* SEM micrographs of longitudinal polished wires with x = 0 and cooled down in furnace. Figs. (a), (b) and (c) correspond to a wire annealed 30 minutes at 700 °C with different enlargements, being (c) a detail of the main phase containing the MgB$_2$ superconductor -circle in (b)-. Fig. (d) Micrograph of a wire annealed 48 hours at 620 °C showing a rich-boron phase (dark contrast).

The existence of finely dispersed Cu-containing phases in the eutectic-like structure as well as larger MgCu$_2$ grains inside the core, may enhance the thermal stability and the mechanical performance of the wires, but it would also result in a reduction of the superconducting volume and therefore in a decrease of the overall $J_c$ of the wires. On the other hand, the presence of relatively large MgCu$_2$ grains inside the core may also affect negatively the superconducting performance of the wires and their homogeneity

In order to analyse the effect of the cooling rate on the microstructure of the samples, two wires with x = 0.2 annealed at 700 °C during 18 minutes, but cooled in the furnace or quenched, were compared. A very similar microstructure was observed in both samples with $MgCu_2$ grains of the same size, although in the quenched wire fewer areas with boron-rich phases appear.

Independently of the annealing temperatures, above and below the Mg melting ($T_m(Mg)$ = 649 °C, [15]) and the treatment times, the same phases are present in the wires. This indicates that some amount of Mg diffuses in Cu, which at the range of annealing temperatures used here 620 - 700 °C and according with the Cu-Mg equilibrium phase diagram [15], may produce liquid phases. Moreover, it has to be noted that at higher annealing temperatures, 780 - 800 °C, there is a large increase of the Mg reactivity with Cu, which results in drastic reductions of the critical current of the wires, even for short annealing times (5 minutes).

*3.2 Transport current at self-field: Influence of heat treatments and initial precursors*

The in-phase component, $\chi'$, of the $\chi_{ac}(T)$ measurements of several wires are displayed in Fig. 3, where for comparative purposes, the results have been scaled using the $|\chi'(5\ K)|$ values. In all samples, $\chi'(T)$ at low enough temperatures reach values close to the corresponding ones to perfect diamagnetism, $\chi'(0\ K) = -1/(1-N)$, $N$ being the demagnetisation factor. For cylinders of high $L/R$ aspect ratio (12 < $L/R$ < 14 in our case) under perpendicular fields $N$ tends to 1/2.

All samples show very similar $T_c$ values, ranging from 38.0 to 38.5 K, but with different transition broadening. Wires fabricated with Mg-excess precursors ($x$ = 0.2) show sharper transitions than stoichiometric ones. Moreover, in both cases, quenched samples have a worse behaviour, that is, wider transitions and slighter lower $T_c$ values. The better transition widths, $\Delta T$, experimentally defined by the temperature range where the ratio $\chi'/\chi'(5\ K)$ changes from 0.9 to 0.1, are $\Delta T$ = 1.5 K. These values are just slightly higher than the ones observed in wires fabricated by similar methods but using harder and non-reactive sheaths, such as iron [5].

The influence of the heat treatment conditions on the transport critical current density values, $J_{c,t}$ may be analysed in Fig. 4(a). These values correspond to 15 K, self-field measurements on a wire fabricated from precursor powders of stoichiometric proportions and cooled down inside the furnace. The error bars of the 700 °C results are typical bounds of measurements on different pieces of the same wire treated under the same conditions.

By annealing at temperatures above $T_m(Mg)$, in the range between 680 and 700 °C, values up to $J_{c,t}$ = 9×10$^8$ A/m$^2$ are obtained for short times (15-30 minutes), while for longer ones, there are important reductions of $J_{c,t}$. On the contrary, for heat treatments at temperatures bellow $T_m(Mg)$ (620 °C, for example) the critical current density increases with the annealing time, and much longer times (48 hours) are needed to reach $J_{c,t}$ values similar to the wires annealed at temperatures above $T_m$. Finally, as pointed out in the above section, for higher annealing temperatures, 800 °C, although the wires are still superconducting, the critical current values decrease drastically, even for very short annealing times (5 minutes) and sharp heating ramps. From these results it is clear that the formation of $MgB_2$ is relatively fast, and short annealing times at relatively low temperatures (700 °C) are sufficient. On the other hand, the observed fast decrease of the critical current density with the annealing time, becomes faster at higher temperatures and is certainly due to the higher diffusivity of Mg in Cu resulting on a higher reactivity.

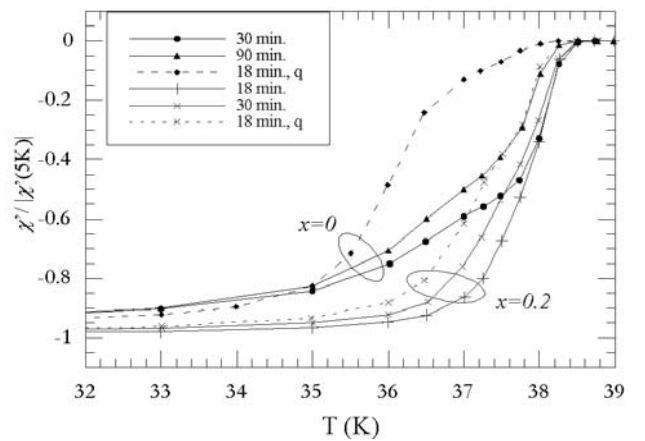

*Figure 3* Results of $\chi'(T)$ under perpendicular ac fields of 1 Hz and field amplitude of 0.1 mT on wires annealed at 700 °C during different times and for two precursor compositions ($x$ = 0 and 0.2). Symbols joined by dashed lines correspond to quenched samples.

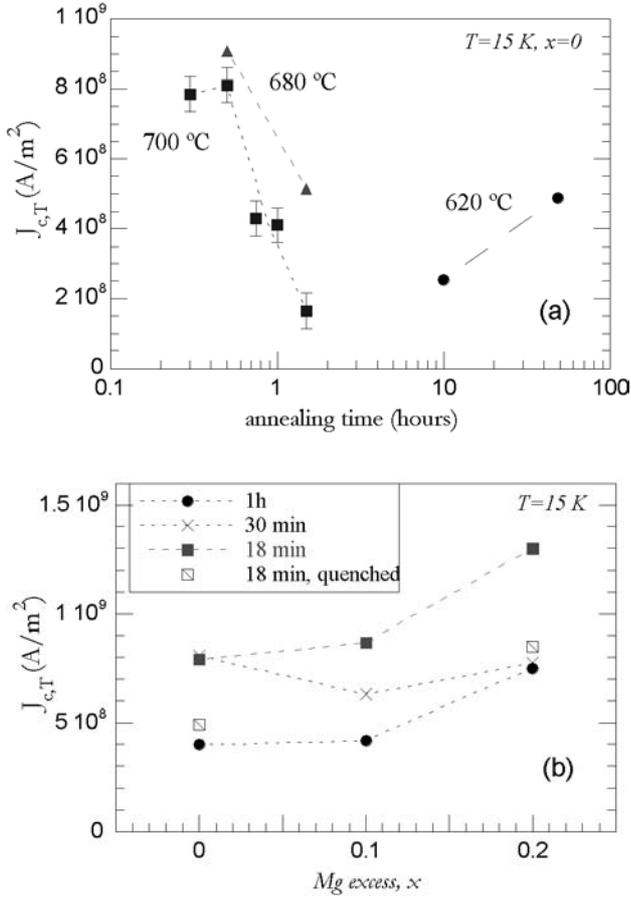

**Figure 4** (a) Transport critical current density, $J_{c,t}$, at 15 K in the self-field of a wire made from stoichiometric precursors for different heat treatments and cooled inside the furnace. (b) $J_{c,t}$ dependence on the initial Mg excess of the precursors, x, for wires annealed at 700 °C and different times, cooled in the furnace (filled symbols and crosses) and quenched (open symbols).

Due to this reactivity with the Cu sheaths, wires from stoichiometric powders develop cores with an overall Mg deficit upon annealing, causing the formation of boron-rich phases and therefore a decrease of the critical current values. In order to analyse this effect, the superconducting properties of wires made from precursor powders with Mg excess have been studied. Fig. 4(b) shows the dependence of the self-field critical current density values at 15 K on the initial Mg excess, $J_{c,t}(x)$, for different annealing times at 700 °C. Results for slow-cooling (full symbols and crosses) and quenched samples (open symbols) show overall increases of the $J_{c,t}$ values with the Mg excess. For the wires with $x = 0.2$, $J_{c,t}$ improves by 50-70% with respect to the ones fabricated from stoichiometric proportions for most annealing conditions, while the properties for the $x = 0.1$ analogue do not differ significantly from those of $x = 0$. Note that although, for all wires, $J_{c,t}$ decreases with the 700 °C annealing times, there are slight differences on the time decays depending on the Mg excess of the precursors. Our results also indicate that quenched samples have considerable lower critical current values than those cooled in the furnace. For self-field conditions, best results of $J_{c,t}(15\ K) = 1.3 \times 10^9$ A/m$^2$, which correspond to a critical current of 700 A, were obtained by annealing 18 minutes at 700 °C and afterwards slow cooling.

The temperature dependences of the self-field transport critical currents of different samples are shown in Fig. 5. The values have been scaled with the measured ones at 15 K. Most samples show a similar behaviour, independently of the annealing temperatures and times, giving an almost linear temperature decay and currents densities close to zero at the 36-37 K range. The wires with poor critical currents (circles) as well as the quenched wires (×), show stronger $J_{c,t}(T)$ dependences and becomes zero at lower temperatures (~35 K), in agreement with the $\chi'(T)$ measurements displayed in Fig. 3. The extrapolation of the above data would give estimates of $J_{c,t} = 2 \times 10^9$ A/m$^2$ at 5 K in the self-field, for the wires with best properties.

All samples show high thermal stability and do not quench during transport measurements, even though these were performed without cryogenic liquid or exchange gas surrounding the sample, being cooled by the good thermal contact of their ends to the cold stage of the cryocooler. For critical current values of 500-700 A, the temperature of the sample, measured by a thermocouple soldered to the sheath, rises between 2 and 3 K during the measurements when the critical current is surpassed up to measured voltages of 10 µV/cm, and has fast recovery to the set-up temperature after the current is switched off.

### 3.3 Magnetic field dependence of $J_c$

The magnetic field dependence of the critical current density have been estimated form the magnetization hysteresis curves measured under perpendicular applied fields ($J_{c,m}$), using critical state models. Fig. 6 shows the results obtained from the width of the hysteresis loop measurements, $\Delta M$ as $J_{c,m} = (3\pi/8)\Delta M/R$, in S.I. units [16], for different wires at 5 and 20 K.

Critical current density values estimated from magnetization measurements, although slightly higher, are in good agreement with those obtained by DC transport (see Fig. 4). Note that on transport self-field measurements, for $J_{c,t} = 10^9$ A/m$^2$, the magnetic field at the surface of the wire core is between 0.2 to 0.3 T.

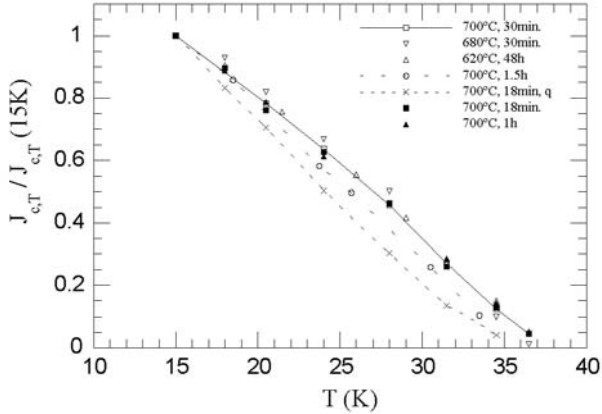

*Figure 5* Temperature dependence of the scaled transport critical current densities in the self-field, $J_{c,t}(T)/J_{c,t}(15\ K)$, for wires annealed in different conditions. 'q' means quenched. Lines are just eye guides.

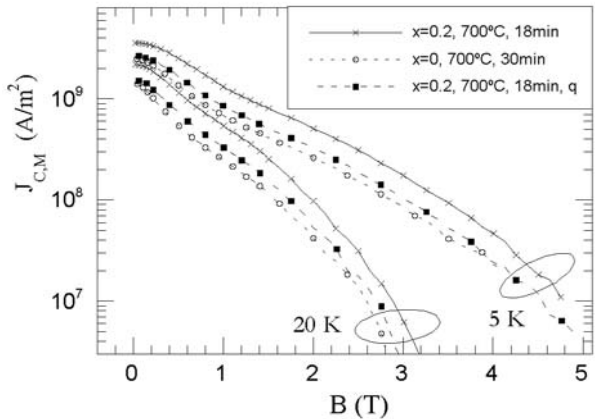

*Figure 6* Inductive critical current density, $J_{c,m}$, estimated from the M-B hysteresis loops at 5 and 20 K for different wires.

Although the $J_{c,m}$ values change among wires, all of them share the same field dependence, indicating that the annealing and cooling conditions or the excess of Mg in the precursor powders essentially do not change the pinning mechanisms of these samples. As it was been already observed [3-6,8-11,13,14,17], this material has a sharp field dependence, in our case with a decrease of $J_{c,m}$ down to $10^8$ A/m$^2$ for magnetic fields of 3.5 T at 5 K and 2 T at 20 K.

Studies reported so far in the literature [17-19], have demonstrated a wide range of $J_c(B)$ behaviours for different polycrystalline MgB$_2$ materials, depending on the geometry and manufacturing techniques. Strong differences among the results on bulk, thin films, wires, tapes, etc, have been observed. A review of such results may be seen for instance in ref. [17].

Among undoped MgB$_2$ materials studied so far, thin films have shown the best superconducting behaviour under magnetic fields, with transport $J_c$ exceeding $10^9$ A/m$^2$ at 4.2 K and 10 T [18]. Although the pinning mechanisms are still not clear, grain boundaries have been suggested to play an important role on polycrystalline materials. In this sense, the typical MgB$_2$ grain size of thin films (less than 10 nm) is much smaller than those of bulk samples, wires and tapes (0.1-1 μm). This difference, together with other defects at the nanoscale range, would explain the larger pinning and therefore the smoother $J_c(B)$ dependencies measured in thin films [18, 19]. Furthermore, other possibilities, such as pinning by oxygen and MgO particles incorporated in thin films have been also proposed.

For wires and tapes, the field decay depends on the metallic sheath, as well as on the precursor powders. The slowest $J_c$ field decays, reported so far, correspond to wires fabricated from reacted MgB$_2$ powders [4, 5]. Again, this better behaviour has been ascribed to vortex pinning along grain boundaries, since typical grain sizes of wires made from unreacted Mg+2B mixtures tend to be larger than those from reacted powders. It must be emphasise that, frequently, optima $J_c$ values at low magnetic fields does not give best results at high magnetic fields [5, 8], because small grains give a larger number of junctions and normally have less current capability at low fields.

The hardness of the sheath as well as the reactivity with the precursors are also key factors. The use of harder and less-reactive metallic sheaths as iron [5], nickel [8] or combinations of different metals such as Nb/Cu/SS or Ta/Cu/SS [4] give rise to different $J_c(B)$ decays depending on the reaction method, *in-situ* or *ex-situ*, as well as on the heat treatment conditions. $J_c$ values higher than $10^8$ A/m$^2$ have been obtained for applied magnetic fields ranging from 3.5 to 6.5 T at 4.2 K [4, 8], and from 2 to 3.5 T at 20 K [5].

The $J_c(B)$ behaviour of the Cu sheathed wires here reported, coincides with results obtained by other groups [19] using the *in-situ* reaction and copper sheaths. These results are of the order, but in the lower limits, of those obtained for wires of harder sheaths as iron, stainless steel and nickel, using the *ex-situ* technique. These differences are due to the *in-situ* reaction it-self and to the reactivity between the Cu sheath and the Mg precursors together with the insufficient hardness of copper. On the other hand, main advantages of these MgB$_2$/Cu wires are certainly related to the good thermal stability given by the high thermal conductivity of the sheath. This allows carrying out transport measurements up to high currents 500-700 A without quenching, which are very likely on samples with others sheaths such as iron, SS, etc [4,5], even without surrounding cryogenic liquid or gas to thermalise the sample. Further efforts are needed to obtain conductors thermally stable and with high performance at fields between 5 to 10 T, both necessary for technical applications.

## 4. Conclusions

We have proved the feasibility of Cu/MgB$_2$ composite monocore wires by the *in-situ* reaction of boron and Mg powders by annealing at temperatures below 700 ºC. A reaction MgCu$_2$ layer of about 20μm thickness surrounding the superconducting core and MgCu$_2$ grains dispersed homogeneously inside the core are formed during the heat treatments.

Due to the reactivity of Mg with the Cu sheath, wires with initial stoichiometric proportions develop an overall Mg deficit during annealing. Therefore, wires prepared from precursor powders with Mg excess, $(1+x)$Mg + 2B, were also analysed. An overall increase of the transport critical current densities, $J_{c,t}$, with the Mg excess on the precursors are observed. For wires with $x = 0.2$, $J_{c,t}$ increases between 50% to 70% the values of wires fabricated from stoichiometric proportions for most annealing conditions. The MgB$_2$ formation is fast, so that shorts annealing times (15 minutes) at 680-700 ºC are sufficient to obtain self-field $J_{c,t}$(15 K) values of the order of $1.3 \times 10^9$ A/m$^2$, corresponding to critical currents, $I_c \sim 700$ A. Longer annealing times reduce $J_c$ values due to the reactivity of Mg and Cu. Our results also indicate that quenched samples have considerable lower critical current values than those cooled down inside the furnace. The field dependence has been also obtained from magnetic measurements, observing a decrease of the $J_{c,m}$ values down to $10^8$ A/m$^2$ for applied magnetic fields of 3.5 T at 5 K and of 2 T at 20 K.

The analysed wires show high thermal stability and do not quench to normal state during transport measurements even without the presence of any cryogenic liquid or gas surrounding the sample. Nevertheless, the porosity of the superconducting core that would affect negatively on the conductors homogeneity, and $J_c(B)$ decays sharper than those obtained on wires prepared by *ex-situ* methods with harder metals, constitute their main disadvantages. Further efforts are needed to obtain conductors which combine both, thermal stabilisation with high performance at high fields (5-10 T) necessary for technical applications.

## Acknowledgements

The financial support of the Spanish CICYT project MAT-2002-04121-C03-02 is acknowledged. E. M. thanks to the *Fondo Social Europeo* and CSIC (I3P program) for her contract. The authors thank Luis Alberto Angurel for useful discussions and José Antonio Gómez for helping in the wires preparation.